\documentclass[fleqn,usenatbib]{mnras}

\usepackage{newtxtext,newtxmath}

\usepackage[T1]{fontenc}

\DeclareRobustCommand{\VAN}[3]{#2}
\let\VANthebibliography\thebibliography
\def\thebibliography{\DeclareRobustCommand{\VAN}[3]{##3}\VANthebibliography}

\usepackage[dvipdfmx]{graphicx}	
\usepackage{amsmath}	
\usepackage{comment}
\usepackage{xcolor}

\newcommand{\aout}{a_\mathrm{out}}
\newcommand{\eout}{e_\mathrm{out}}
\newcommand{\ainn}{a_\mathrm{inn}}
\newcommand{\einn}{e_\mathrm{inn}}

\newcommand{\kms}{{\,\rm km/s}}
\newcommand{\yr}{{\,\rm yr}}
\newcommand{\au}{{\,\rm au}}
\newcommand{\gyr}{{\,\rm Gyr}}
\newcommand{\myr}{{\,\rm Myr}}
\newcommand{\msun}{{\,\rm M_\odot}}
\newcommand{\rsun}{{\,\rm R_\odot}}
\newcommand{\zsun}{{\,\rm Z_\odot}}

\newcommand{\chip}{{\chi_\mathrm{p}}}
\newcommand{\chieff}{{\chi_\mathrm{eff}}}

\newcommand{\rev}[1]{{#1}}

\newcommand{\OCZ}[1]{\texttt{OC\_Z#1}}
\newcommand{\GCZ}[1]{\texttt{GC\_Z#1}}

\title[Spin misalignment in black hole binaries]{Spin misalignment of black hole binaries from young star clusters: implications for the origin of gravitational waves events}

\author[A. A. Trani et al.]{
	A. A. Trani,$^{1,2}$\thanks{E-mail: aatrani@gmail.com}
	A. Tanikawa,$^{1}$
	M. S. Fujii,$^{3}$
	N.W.C. Leigh,$^{4,5}$
	J. Kumamoto,$^{3}$
	\\
	$^{1}$Department of Earth Science and Astronomy, College of Arts and Sciences, The University of Tokyo, 3-8-1 Komaba, Meguro-ku, Tokyo 153-8902, Japan \\
	$^{2}$Okinawa Institute of Science and Technology, 1919-1 Tancha, Onna-son, Okinawa 904-0495, Japan \\
	$^{3}$Department of Astronomy, Graduate School of Science, The University of Tokyo, 7-3-1 Hongo, Bunkyo-ku, Tokyo 113-0033, Japan\\
	$^{4}$Departamento de Astronom\'ia, Facultad de Ciencias F\'isicas y Matem\'aticas, Universidad de Concepci\'on, Concepci\'on, Chile \\
    $^{5}$Department of Astrophysics, American Museum of Natural History, New York, NY 10024, USA\\
}
\date{Accepted XXX. Received YYY; in original form ZZZ}

\pubyear{2021}

\begin{document}
\label{firstpage}
\pagerange{\pageref{firstpage}--\pageref{lastpage}}
\maketitle

\begin{abstract}
	Recent studies indicate that the progenitors of merging black hole (BH) binaries from young star clusters can undergo a common envelope phase just like isolated binaries. If the stars emerge from the common envelope as naked cores, tidal interactions can efficiently synchronize their spins before they collapse into BHs. Contrary to the isolated case, these binary BHs can also undergo dynamical interactions with other BHs in the cluster before merging. The interactions can tilt the binary orbital plane, leading to spin-orbit misalignment.
	We estimate the spin properties of merging binary BHs undergoing this scenario by combining up-to-date binary population synthesis and accurate few-body simulations. We show that post-common envelope binary BHs are likely to undergo only a single encounter, due to the high binary recoil velocity and short coalescence times.
	Adopting conservative limits on the binary-single encounter rates, we obtain a local BH merger rate density of ${\sim} 6.6 \yr^{-1} \,\rm Gpc^{-3}$.
	Assuming low (${\lesssim}0.2$) natal BH spins, this scenario \rev{reproduces the trends in the distributions of effective spin $\chi_\mathrm{eff}$ and precession parameters $\chip$ inferred from GWTC-2, including the peaks at $(\chi_\mathrm{eff}, \chi_\mathrm{p}) \sim (0.1, 0.2)$ and the tail at negative $\chi_\mathrm{eff}$ values}.
\end{abstract}




\section{Introduction}

The origins of gravitational wave (GW) events from merging black hole (BH) binaries is still disputed. The numerous formation pathways that have been proposed so far can be organized into two broad categories: dynamically interacting binaries in dense environments and isolated binaries evolving due to stellar and binary evolution in the field. 

The scenarios belonging to the first category include binaries dynamically assembled in globular clusters \citep[GCs, e.g. ][]{sigurdsson1995,downing2010,tanikawa2013a,bae2014,leigh2014b,rodriguez2016,askar2017,samsing2018,arcasedda2018,hong2018,antonini2020b} and other dense stellar environments \citep[open, young and nuclear star clusters, e.g.][]{portegieszwart2000,ziosi2014,fujii2017,banerjee2017,banerjee2017a,petrovich2017,leigh2018,rodriguez2018b,banerjee2018,rastello2018,micheaely2019,dicarlo2019,hong2020,rastello2020,mapelli2020,banerjee2021a}, mergers mediated by dynamical interactions with a supermassive BH or an AGN disk \citep[e.g. ][]{,antonini2012,antonini2016,lei16a,vanlandingham2016,stone2016,bartos2017,hamers2018,mckernan2018,hoa18,yang2019,trani19b,mckernan2020,tagawa2020,arcasedda2020}, and mergers in multiple stellar systems \citep[e.g.][]{antonini2014,silsbee2017,antonini2017,rodriguez2018,hamers2020,martinez2020,leigh2020}.

Conversely, in the field scenario, a primordial stellar binary becomes a merging binary BH via stellar interactions, either via common envelope evolution in Population I/II \citep[e.g.][]{tutukov1973,bethe1998,dominik2012,dominik2013,belczynski2002,belczynski2008,belczynski2010,belczynski2016,belczynski2020,mennekens2014,stevenson2017,spera2015,spera2019,krukow2018,eldridge2016,eldridge2019,mapelli2018,mapelli2017,mapelli2019,neijssel2019,bavera2020,vignagomez2020,banerjee2021a} and Population III stars \citep[e.g.][]{belczynski2004,kinugawa2014,inayoshi2017,kinugawa2016,tanikawa2020c,tanikawa2021}, or via chemically homogeneous evolution \citep[e.g.][]{mink2016,mandel2016,marchant2016,riley2020,buisson2020}.

The total number of events announced by Ligo-Virgo-KAGRA (hereafter LVK) collaboration amounts to 50: 11 from the O1/O2 observing runs \citep{gwtc-1}, and 39 from the first half of the O3 observing run \citep{gwtc2020a}. While we are currently far from being able to discriminate the formation pathways of individual events, as the number of detections increases it will be possible to infer the origin of GW events in a statistical sense, i.e. from the ensemble properties of the entire population \citep[e.g.][]{zevin2017,fishbach2018,bouffanais2019,zevin2020,bouffanais2021}. In this work we focus on one of the most informative properties of GW progenitors: the spin parameter.

Recently, an interplay between the two main formation scenarios has been proposed. \citet{kumamoto2019} showed that, in young star clusters, binary BHs formed via common envelope evolution of dynamically assembled \textit{main sequence} binaries.  These systems contribute to the merger rate more than dynamically assembled binary BHs. These binaries undergo a common envelope phase, like in the isolated channel, but they also undergo dynamical interactions before and after collapsing to BHs. The common envelope phase might be even triggered by such dynamical interactions. Similar evolutionary pathways were also found by \citet{dicarlo2020b}.

Binaries undergoing this channel experience a common envelope phase, which dramatically decreases the binary separation. Consequently, tidal interactions in the post-common envelope phase can efficiently spin up the naked stellar cores \citep[][]{kushnir2016,piran2020}, aligning their spin vectors with the orbital angular momentum. However, this alignment might be lost during a subsequent dynamical encounter in the cluster.

In this work we investigate the effective spin distributions of merging binary BHs from this hybrid pathway. Specifically, we consider post-common envelope binary BHs that undergo a single dynamical encounter in stellar clusters. In Section~\ref{sec:ic} we discuss our numerical setup and initial assumptions. Section~\ref{sec:res} presents the outcome properties of the three-body encounters, while in Section~\ref{sec:encrates} we discuss the encounter rates in cluster environments. In Section~\ref{sec:disc} we calculate the merger properties of post-encounter binaries and compare them with the recent GWTC-2 data, including the local merger rate density (Section~\ref{sec:locmerg}), the effective spin distributions per stellar metallicity (Figure~\ref{fig:Xcum}) and the differential merger rate density at the current epoch (Figure~\ref{fig:Xrates}).


\section{Numerical setup}\label{sec:ic}

We perform direct N-body simulations of binary-single encounters using \textsc{tsunami} (A.A. Trani et al., in preparation). \textsc{tsunami} employs a combination of numerical techniques to ensure excellent accuracy over a wide dynamical range. First, we solve the equations of motion derived from a time-transformed Hamiltionian. Specifically, here we use a 2nd order Leapfrog method from the regularized Logarithmic Hamitonian of \citet[][see also \citealt{preto1999}]{mik99a}. We then increase the accuracy of the integrator using the Bulirsch-Stoer extrapolation \citep{stoer1980}. Finally, the equations of motion are solved in a relative chain coordinate system, as in \citep{mik93}, in order to reduce round-off errors due to small interparticle distances far from the origin in the classical center of mass coordinates. We also include post-Newtonian corrections to the equations of motion, specifically the 1PN, the 2PN and 2.5PN terms \citep{blanchet2014}.

The initial setup is as follows: the binary center-of-mass and the single lie on a hyperbolic \rev{(i.e. unbound)} orbit with \rev{negative} semimajor axis $\aout$ and eccentricity $\eout$. The mass of the binary members and the single are $m_1$, $m_2$ and $m_3$, respectively. The inner binary is on a \rev{(bound) elliptic} orbit with semimajor axis $\ainn$ and eccentricity $\einn$, and has a random orientation uniform on a sphere. 

The orbital parameters and masses of the inner binary are taken from population synthesis simulations using a modified version of the \textsc{bse} code \citep{hur02}.
The initial conditions of the population synthesis simulations are the following. The initial mass function (IMF) of primary stars follows the \citep{kro01} IMF truncated from $10 \msun$ to $150 \msun$. The secondary-to-primary mass ratios are drawn from a flat distribution between $0$ and $1$, imposing a minimum mass of $10 \msun$ for the secondary star. The binary semi-major axes have a flat distribution in logarithmic scale from $1 \rsun$ to $10^6 \rsun$ \citep{andrews2017}, and their eccentricities follow a thermal distribution \citep{jeans1919}. The initial pericenter distances are set to be large enough to avoid the onset of Roche-lobe overflow. \rev{Note that by simulating the binary stellar evolution in isolation we are neglecting common envelope events triggered by dynamical encounters, and therefore we might be underestimating the number of post-common envelope binaries.}

We evolve 3 sets of different metallicities: $1$, $0.1$, and $0.01 \zsun$ \rev{(where $\zsun = 0.02$)}, running $10^5$ realizations for each set. The stellar wind models and binary interaction model are the same as those in \citet[][see also \citealt{giacobbo2018}]{tanikawa2020c}. We do not take into account BH natal kicks. Since there is no mechanism which can tilt BH spins from the orbital planes, the BH spins will be parallel to the orbital planes due to tidal synchronization \citep{hut81}.

We select those binaries which are likely to have spinning BHs. Specifically, we select only those binaries whose separation is small enough to allow the tidal spin-up of the stars, after the binary exits the common envelope phase as a double Wolf-Rayet star. \rev{As shown by \citet[][see also \citealt{safarzadeh2020}]{kushnir2016}, the value of the  synchronization spin scales as $\chi \propto d^{-3/2}$, where $d$ is the binary separation. Given that the GW coalescence timescale scales as $t_\mathrm{gw} \propto d^4$, it follows that $\chi \propto t_\mathrm{gw} ^{-3/8}$: the shorter the coalescence time, the higher the dimensioness spin}. Therefore, we select only those binaries that have a GW merging timescale of ${\lesssim}200 \myr$ (before turning into BHs) and we exclude those binaries whose GW timescales are too small (${<}50 \myr$) to allow for encounters with other stars before their mergers.

The outer orbit is determined in the following way. The velocity at infinity $v_\infty$ of the outer orbit is drawn from a Maxwellian distribution with $\sigma_\infty$ dispersion. We consider two main environments in which the three-body encounters take place: open clusters (OCs), which have low velocity dispersions, and massive clusters with high velocity dispersions, such as globular and young massive star clusters. We set $\sigma_\infty$ to $1\kms$ and $20\kms$ for OC and GC simulations, respectively. Note that this represents the 3D velocity dispersion, rather than the commonly reported line-of-sight velocity dispersion, which is a factor of $\sqrt{3}$ smaller \citep{illingworth1976,harris1996,por10}. \rev{Given $v_\infty$, the semimajor axis of the outer hyperbolic orbit is $\aout = - G m_\mathrm{tot} / v_\infty^2$}, where $m_\mathrm{tot} = m_1 + m_2 + m_3$ is the total mass of the bodies. \rev{The eccentricity $\eout$ is instead calculated from the pericenter distance $p_\mathrm{out} = \aout (1-\eout)$, which is drawn from} a distribution uniform in $p_\mathrm{out}^2$ between $0$ and $2\ainn$. This is chosen so as to minimize the number of flyby interactions, since numerical experiments have shown that the cross section for resonant encounters drops for $p_\mathrm{out}>2\ainn$ \citep[e.g.][]{hut83a,hut83b,hut93}. 

\begin{figure}
	\includegraphics[width=\columnwidth]{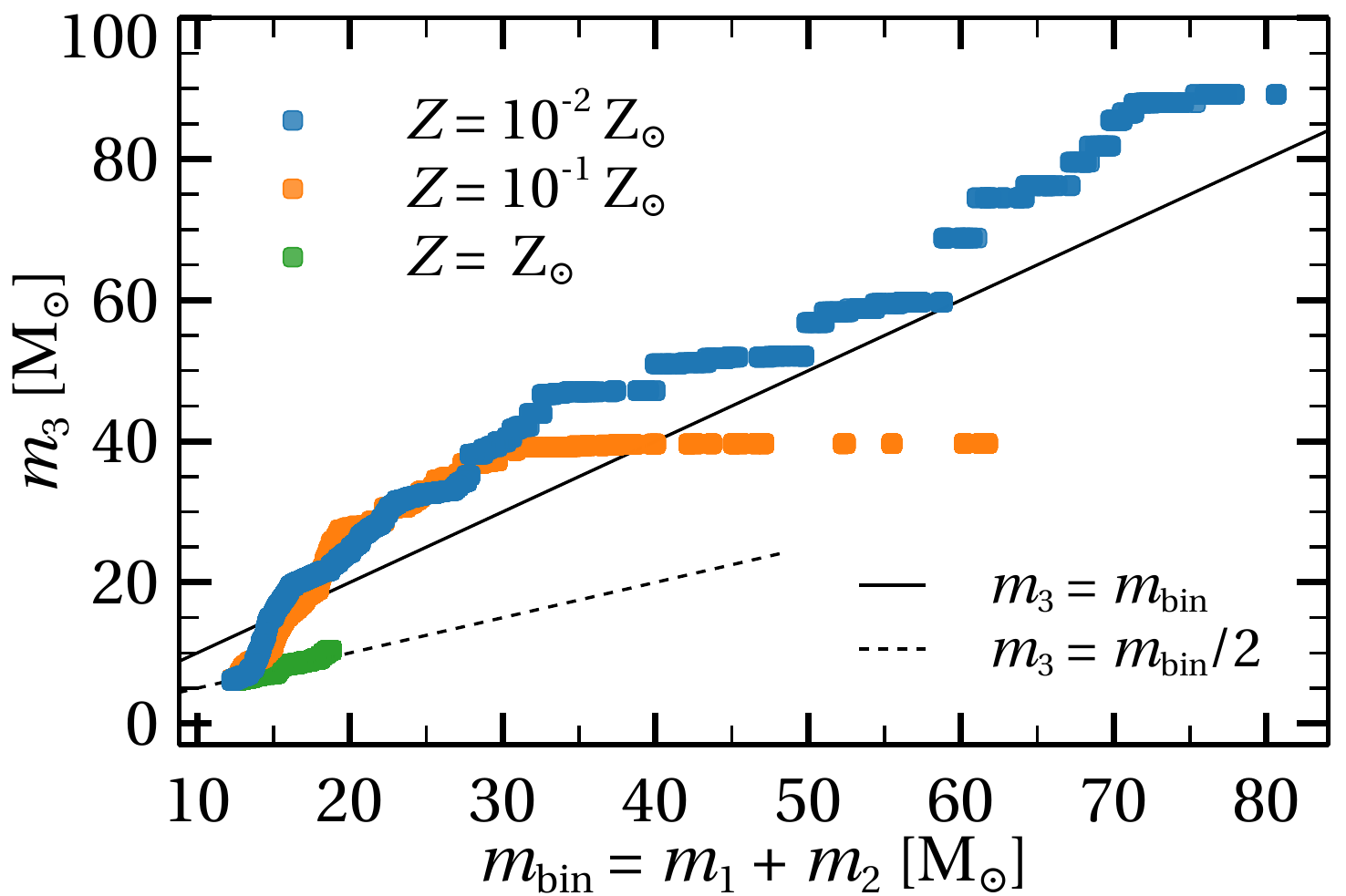}
	\caption{Mass of the single body versus mass of the binary for each binary-single simulation. Blue squares: $Z=10^{-2}\zsun$. Orange squares: $Z=10^{-1}\zsun$. Green squares: $Z=\zsun$. The black dashed lines denotes $m_1 + m_2 = m_3$. At low metallicities, the single BHs are more massive than the BHs in binaries, because they dot not lose mass via binary interactions.}
	\label{fig:mA_mB}
\end{figure}

The mass of the third body is drawn from the stellar population synthesis code \textsc{sse}, in accordance with the \textsc{bse} simulations used for the inner binaries. It is well known that massive stars and binaries sink to the cores of star clusters faster than lighter stars, due to mass segregation and energy equipartition. \citep[e.g.][]{giersz1996,gur04,khalisi2007,goswami2012,tanikawa2012,tanikawa2013,tra14,fujii2014,spera2016,webb2016,pavlik2020}. We mimic this effect by pairing the most massive binaries to the most massive single body, as shown in Figure~\ref{fig:mA_mB}.
We combine 3 choices of stellar metallicity with 2 choices of velocity dispersion,  for a total of 6 sets of simulations. Each set comprises of $10^5$ realizations, enough to sample the initial parameter space. Table~\ref{tab:ic} summarizes the initial conditions for each set.

The initial binary-single distance is set to $100 \ainn$, which is large enough so that the initial binary is unperturbed by the single. The simulations are run until either the three-body interaction is complete (with an outbound binary-single on a hyperbolic orbit), or there is a merger.

\begin{table}
	\centering
	\caption{Initial conditions.}
	\label{tab:ic}
	\begin{tabular}{lcc} 
		\hline
		Set name & $\sigma_\infty$ [$\kms$] & $Z$ [Z$_\odot$] \\
		\hline
		\OCZ2 & 1 & 0.01 \\
		\GCZ2 & 20 & 0.01 \\
		\OCZ1 & 1 & 0.1 \\
		\GCZ1 & 20 & 0.1 \\
		\OCZ0 & 1 & 1 \\
		\GCZ0 & 20 & 1 \\
		\hline
	\end{tabular} \\
	{\scriptsize $\sigma_\infty$: dispersion of the Maxwellian distribution for $v_\infty$ \\
		$Z$: metallicity of the population synthesis simulations}
\end{table}

\section{Encounter outcomes}\label{sec:res}

A three-body encounter can be represented as a succession of strong chaotic interactions, wherein all three-bodies exchange energy, and long, regular excursions in which the single is ejected from the binary and forms a temporary hierarchical bound system \citep[e.g.][]{stone2019}. Eventually, assuming all point particles, the single will achieve a velocity above the escape velocity, and the encounter will end with an unbound binary-single. In the context of three-body interactions in the cores of stellar clusters, this process usually leads the binary semimajor axis to shrink, and it is therefore known as binary hardening \citep{heg75}.

After the encounter is concluded, the orientation of the outgoing binary might be different than the initial one. This will cause spin-angle misalignment if initially the BH spins were aligned with the orbit. We measure the tilt angle $\Delta i$ between the initial binary plane and the final one, and check if one of the binary members was exchanged with the single BH. Hereafter, we refer to binaries that underwent an exchange as ``exchanged binaries'', while the binaries that did not are termed ``original binaries''.

\begin{figure}
	\centering
	\includegraphics[width=1\columnwidth]{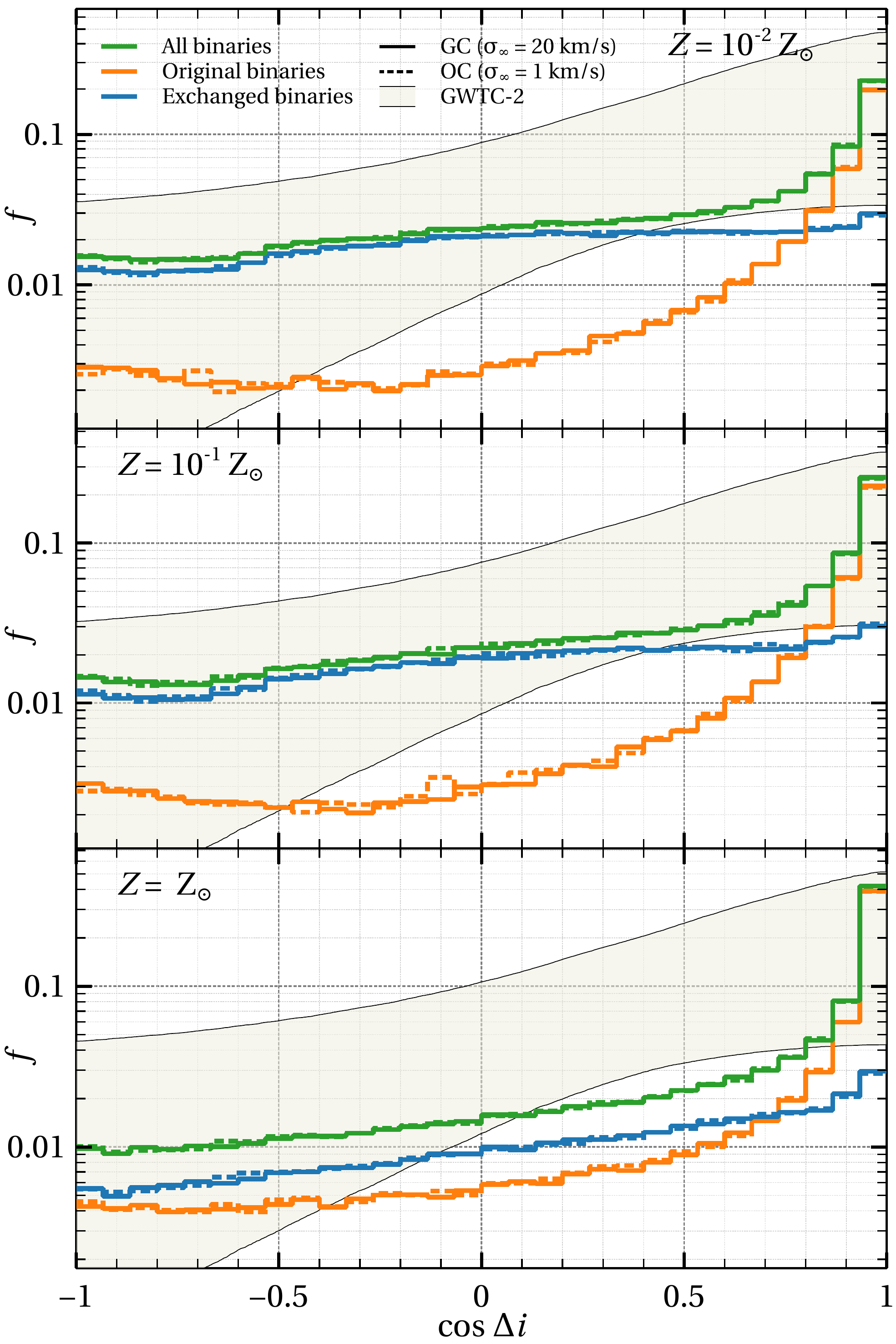}
	\caption{Distribution of the misalignment angle $\Delta i$ for each of our simulation sets. Solid lines indicate GC initial conditions with $\sigma_\infty = 20 \kms$, while the dotted lines indicate OC initial conditions with $\sigma_\infty = 1 \kms$. Green lines: all binaries. Orange lines: original binaries. Blue lines: exchanged binaries.
	Top panel: $Z=10^{-2}\zsun$. Middle panel: $Z=10^{-1}\zsun$. Bottom panel: $Z=\zsun$. The gray shaded area shows the 90\% confidence interval of the  distribution reconstructed from GWTC-2 \citep[default spin model from][]{gwtc2020b}.}
	\label{fig:dangletot}
\end{figure}

Figure~\ref{fig:dangletot} shows the distributions of $\cos{\Delta i}$ for all our sets. The distributions are essentially the same regardless of the initial velocity dispersion, indicating that encounters in OCs and GCs will lead to very similar tilt angles. This is because the initial binaries are very tight, and the single BH very massive, so that regardless of $\sigma_\infty$ we are in the regime of hard binary scatterings \citep{heg75,hut83b,hut1984}. 
The tilt angle distribution of original binaries is strongly peaked at $\cos{\Delta i}=1$, indicating that most original binaries retain their initial orientation. 
The peak at $\cos{\Delta i}=1$ becomes more pronounced with increasing metallicity: about 15\%, 18\% and 34\% of the binaries get misaligned by less than 15$^\circ$ at $Z=0.01$, $0.1$ and $1\zsun$, respectively. The tilt angle distribution of exchanged binaries is moderately flat, except at $Z=1\zsun$ where it becomes more strongly peaked at $\Delta i=0^\circ$

\begin{figure}
	\centering
	\includegraphics[width=1\columnwidth]{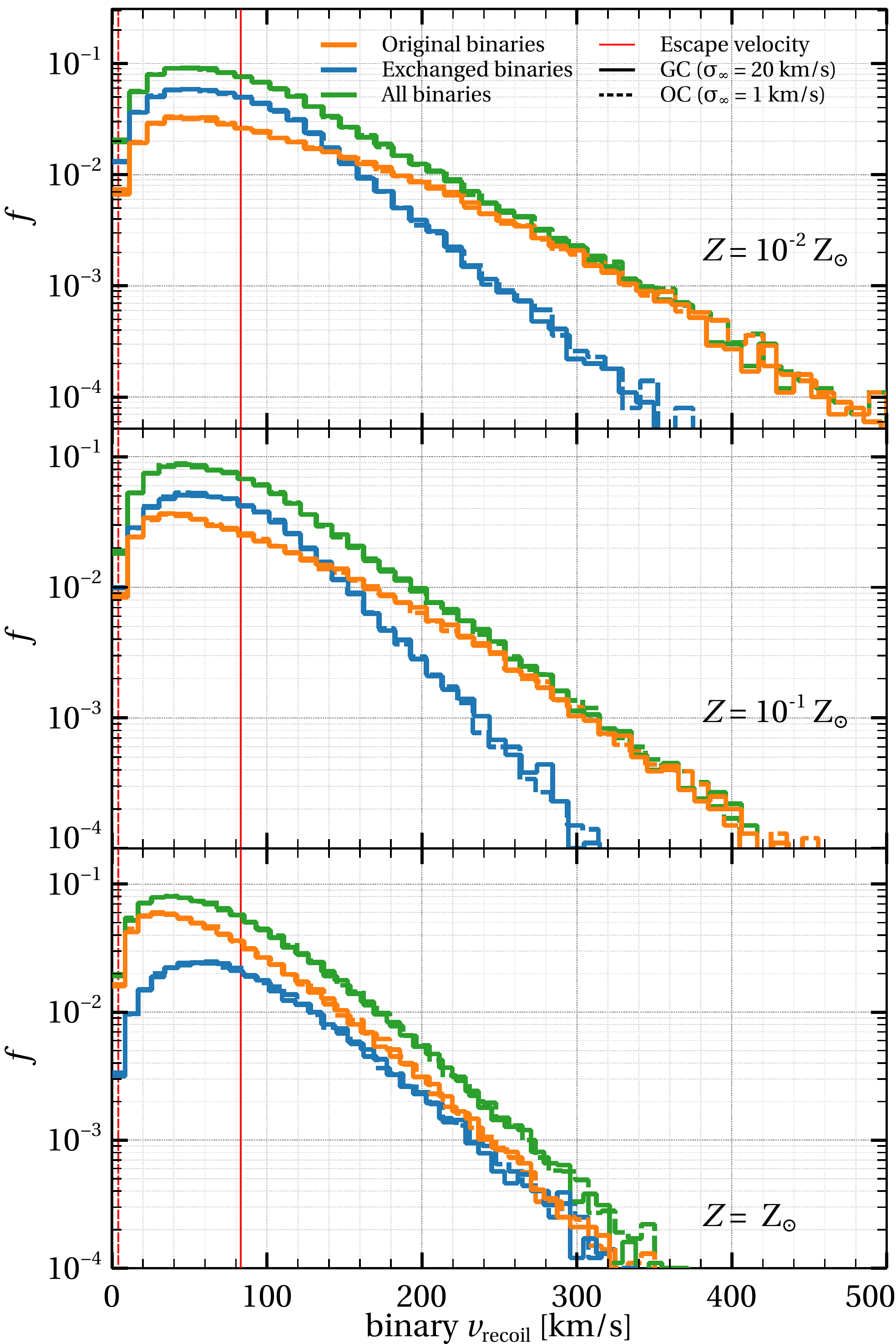}
	\caption{Distribution of the binary recoil velocity for each of our sets. Solid lines indicate GC initial conditions with $\sigma_\infty = 20 \kms$, while dotted lines indicate OC initial conditions with $\sigma_\infty = 1 \kms$. Green lines: all binaries. Orange lines: original binaries. Blue lines: exchanged binaries. The vertical red lines indicates the cluster escape velocity for GC and OC cases.
		Top panel: $Z=10^{-2}\zsun$. Middle panel: $Z=10^{-1}\zsun$. Bottom panel: $Z=\zsun$}
	\label{fig:vescplot}
\end{figure}

In Figure~\ref{fig:vescplot} we show the distributions of binary recoil velocity for all our sets. As for the $\cos{\Delta i}$ distribution, there is little dispersion between OC and GC cases. However, the escape velocity from OCs is much lower than for GCs. As a conservative choice, we set the cluster escape velocity to $v^\mathrm{cluster}_\mathrm{esc} = 4 \sigma_\infty$ \rev{(i.e. twice as high as it would be from the virial relation)}, which amounts to $80 \kms$ and $4 \kms$ for the GC and OC sets, respectively. This results in the majority (${>}97\%$) of binaries escaping in the OC sets, while only about $40\%$ of the binaries escape in the GC sets.

The kick distribution shifts towards lower velocities at higher metallicity. This is a consequence of the lower BH masses at high metallicity, which makes three-body interactions less energetic and reduces the overall impulse imparted on any given particle, such that higher ejection velocities are harder to achieve.

Except at $Z=1\zsun$, the high-velocity tail of the recoil distribution is dominated by the original binaries, while the low-end is dominated by exchanged binaries. The reason for this is that during exchanges, a light BH is replaced by a more massive one, so that the exchanged binary is more massive than the original. The binary recoil kick is proportional to the mass of the single, which makes the recoil kick of the original binaries higher than that of the exchanged ones, even if the kick magnitude is generally higher in exchanges.

Table~\ref{tab:out} summarizes the outcome fractions of the simulations. In the \GCZ2 set, ${\sim}40\%$ of the final binaries are original binaries. However, since original binaries have larger recoil velocities, $60\%$ of them escape from the cluster, compared to only $46\%$ of the exchanged binaries.

\begin{table}
	\centering
	\caption{Fractional outcomes of the simulation sets.}
	\label{tab:out}
	\begin{tabular}{lcccccc} 
		\hline
		Set name & $f_\mathrm{ori}$ & $f_\mathrm{ex}$ & $f_\mathrm{esc}$ & $f_\mathrm{ori,esc}$  & $f_\mathrm{ex,esc}$ & $f_\mathrm{merg}$ \\
		\hline
		\OCZ2 & 0.410 & 0.578 & 0.986 & 0.409 & 0.577 & 0.011 \\
		\GCZ2 & 0.409 & 0.579 & 0.474 & 0.227 & 0.247 & 0.011 \\
		\OCZ1 & 0.439 & 0.551 & 0.987 & 0.438 & 0.549 & 0.010 \\
		\GCZ1 & 0.443 & 0.546 & 0.441 & 0.216 & 0.225 & 0.010 \\
		\OCZ0 & 0.661 & 0.320 & 0.977 & 0.657 & 0.320 & 0.019 \\
		\GCZ0 & 0.660 & 0.322 & 0.367 & 0.221 & 0.146 & 0.018 \\
		\hline
	\end{tabular} \\
	{\scriptsize $f_\mathrm{ori}$: fraction of original binaries;  
		$f_\mathrm{ex}$: fraction of exchanged binaries; 
		$f_\mathrm{esc}$: fraction of escaped binaries; 
		$f_\mathrm{ori,ex}$: fraction of original binaries that achieve the escape speed;
		$f_\mathrm{ex,esc}$: fraction of exchanged binaries that achieve the escape speed;
		$f_\mathrm{merg}$: fraction of mergers during the encounter}
\end{table}

\begin{figure}
	\centering
	\includegraphics[width=1\columnwidth]{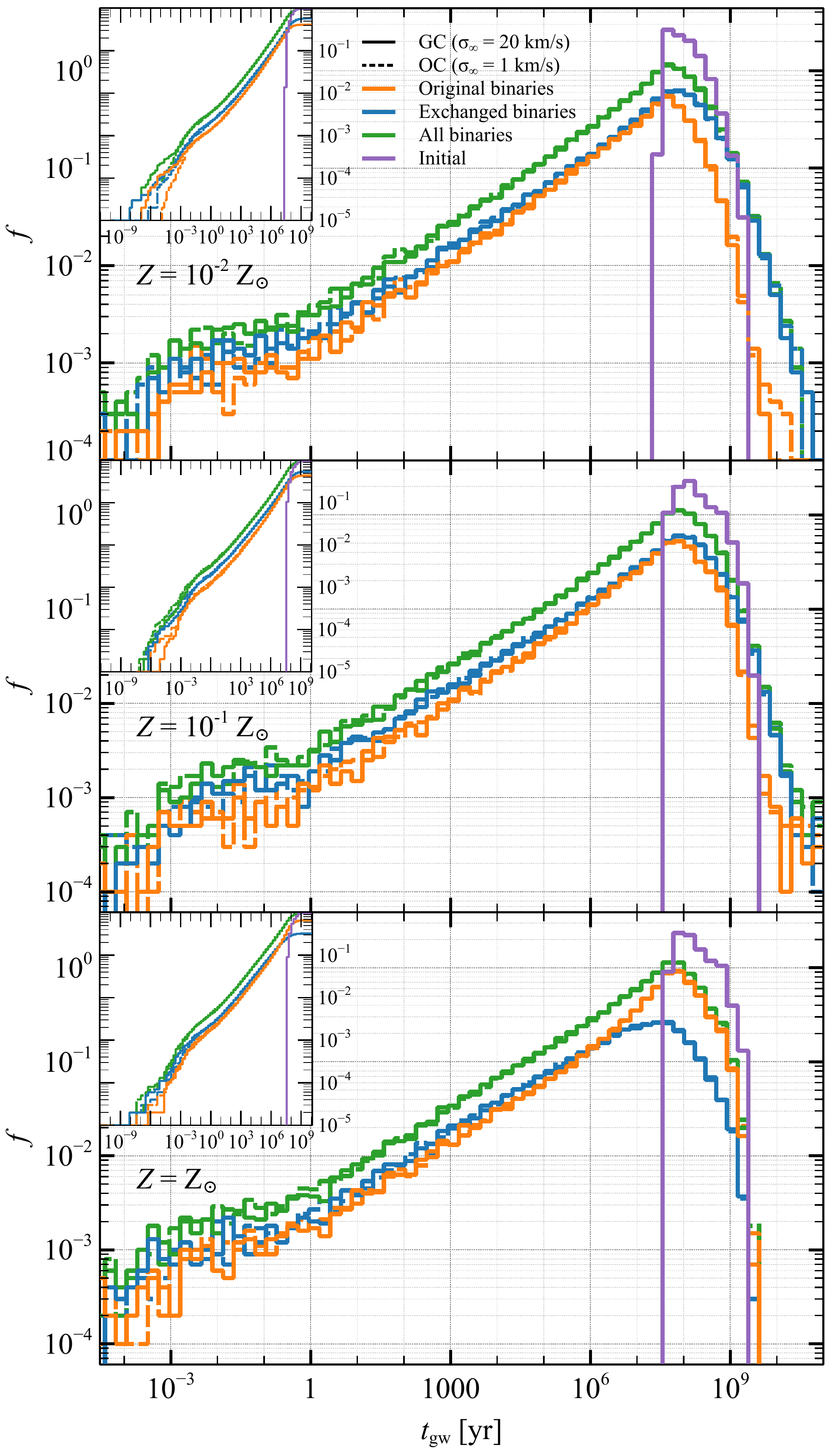}
	\caption{Distribution of the GW-coalescence time for each of our sets. Solid lines indicate GC initial conditions with $\sigma_\infty = 20 \kms$, while dotted lines indicate OC initial conditions with $\sigma_\infty = 1 \kms$. Green lines: all binaries. Orange lines: original binaries. Blue lines: exchanged binaries. Purple line: initial distribution.
		Top panel: $Z=10^{-2}\zsun$. Middle panel: $Z=10^{-1}\zsun$. Bottom panel: $Z=\zsun$. The inset shows the cumulative distribution. }
	\label{fig:tgw}
\end{figure}

The distribution of the GW coalescence times $t_\mathrm{gw}$ is shown in Figure~\ref{fig:tgw}. \rev{We calculate the coalescence time using the following expression from \citet{peters64}:
\begin{equation}\label{eq:delay}
	t_\mathrm{gw} = \frac{15 c^5}{304 G^3} \frac{a^4}{(m_1+m_2)\,m_1\,m_2}\,f(e)
\end{equation}
where $f(e)$ is a factor taking into account the orbital eccentricity that we evaluate numerically as:
\begin{equation}\label{eq:fe}
f(e) = \frac{(1-e^2)^4}{e^{\frac{48}{19}}(e^2 + \frac{304}{121})^{\frac{3480}{2299}}} \int^e_0 \frac{x^{\frac{29}{19}} (1 + \frac{121}{324}x^2)^{\frac{1181}{2299}}}{(1-x^2)^{3/2}} dx
\end{equation}

Figure~\ref{fig:tgw} includes all the binaries}, whether they escaped from the cluster or not, because the encounter rate of non-escaped binaries is small compared to their coalescence time, so that it is unlikely that they undergo a second encounter even if they do not escape from the cluster (see Section~\ref{sec:encrates}). For all metallicities, the initial distribution peaks at ${\approx} 100\myr$ and ranges from $40 \myr$ to $10 \gyr$. \rev{This is the result of our selection of the initial binary sample: we included only those BH binaries whose progenitors have a coalescence time of ${<}200 \myr$ after the common envelope phase, so that they have a chance to be spun-up by tidal interactions.}

The peak coalescence time in the final distribution for all binaries becomes slightly shorter, at ${\approx}10\myr$ after the encounter. 
At $Z=\zsun$, this peak is dominated by exchanged binaries: most of the original binaries have a much shorter coalescence time, peaked at ${\approx}1\myr$. 
At higher metallicity, the discrepancy in coalescence time between exchanged and original binaries lessens. 

The reason for this is that at low metallicity, the original binaries have a shorter semimajor axis than the exchanged ones. Since at low metallicity the initial single star can be twice more massive than the binary (see Figure~\ref{fig:mA_mB}), the binary needs to harden more in order to eject the single. On the other hand, if the initial single ejects one of the lower-mass binary components, the binary can maintain a larger separation. 

Another way to phrase it is that the outcome distribution of binary binding energies $E_\mathrm{b} = G m_\mathrm{1} m_\mathrm{2} / 2a$ of a three-body encounter is the same whether the final binary is the original or exchanged. Therefore, at the same $E_\mathrm{b}$, a higher mass product $m_\mathrm{1} m_\mathrm{2}$ translates into a larger semimajor axis $a$, and vice versa \citep{valtonen2005}.
However, at short coalescence times the distribution has a similar trend for both exchanged and original binaries. 

\section{Encounter rate estimates}\label{sec:encrates}
The merger rate density for these kinds of events depends linearly on the encounter rate between the compact binary BHs and single BHs. Here we estimate and discuss the encounter rate for the three-body encounters considered in this work. The encounter rate of binary-single encounters $\Gamma^{2+1}_\mathrm{enc}$ can be expressed as \citep{leigh2011}:
\begin{equation}\label{eq:encrates}
\Gamma^{2+1}_\mathrm{enc} = N_{\rm sin} n_{\rm bin} \frac{3\pi G m_\mathrm{tot} R_\mathrm{enc}}{\sigma_\infty} = f_{\rm bin} f_{\rm sin} \, N n \, \frac{3\pi G m_\mathrm{tot} R_\mathrm{enc}}{\sigma_\infty}
\end{equation}
where $n$ is the stellar number density, $N$ is the number of stars, $\sigma_\infty$ is the velocity dispersion, $m_\mathrm{tot} = m_\mathrm{bin} + m_\mathrm{3}$ is the mass of the binary plus single, $f_{\rm bin}$ and $f_{\rm sin}$ are the binary and single fractions, and $R_\mathrm{enc}$ is the distance below which the binary-single can undergo a chaotic three-body encounter, which we conservatively set to $2\ainn$. Here we have assumed that the cross section is dominated by gravitational focusing, so that the Safronov number $\Theta = G m_\mathrm{tot}/ R_\mathrm{enc}\sigma_\infty^2 \gg 1$ \citep{galacticdynamics}. This latter condition is always satisfied given the compactness of the binary.

The velocity dispersion depends on cluster size and mass via the virial relation $\sigma_\infty^2 = 0.45 G M_{\rm cl} / r_\mathrm{h}$, where $M_{\rm cl}$ is the cluster mass and $r_\mathrm{h}$ is the half-mass radius. We can eliminate the dependence on $r_\mathrm{h}$ via the Marks-Kroupa relation \citep{marks2012,leigh13,leigh15}, which relates cluster mass at birth to its half-mass radius. Given an average stellar mass $\langle m\rangle$, we can re-express $N = M_{\rm cl}/\langle m\rangle$ and $n = \rho/\langle m\rangle$, we obtain an encounter rate that depends only on cluster density, mass and the relative fractions of objects undergoing the encounter:

\begin{equation}
\begin{aligned}
	\Gamma^{2+1}_\mathrm{enc}(M_\mathrm{cl}) \simeq 30 \myr^{-1} & f_{\rm bin} f_{\rm sin} \left(\frac{0.5 \msun}{\langle m \rangle}\right)^2 \left(\frac{\rho}{10^5 \msun{\rm pc^{-3}}}\right) \\ &  \left(\frac{m_\mathrm{tot}}{60 \msun}\right) \left(\frac{R_\mathrm{enc}}{0.072 \au}\right) \left(\frac{M_\mathrm{cl}}{10^3 \msun}\right)^{0.565} 
\end{aligned}
\end{equation}

Here we have adopted the median values for our three-body simulations ($m_\mathrm{tot}\simeq60\msun$, $\ainn \simeq 0.036 \au$), and assumed that encounters occur mainly during the core collapse phase in stellar clusters, when the core density can peak at $>10^4$--$10^5 \msun/{\rm pc}^3$. 

\rev{In the above equation, $f_\mathrm{bin} \equiv f_{\rm CBBH}$ is the fraction of post-common envelope binaries representative of our pre-encounter binary sample. This fraction can be estimated from our binary population synthesis simulations as the ratio between the number of selected binaries and the initial number of binary realizations}. This fraction depends on the metallicity, and amounts to $7.6 \times 10^{-5}$, $1.02 \times 10^{-3}$ and $2.17 \times 10^{-3}$ for $Z = 1, 0.1$ and $0.01 \zsun$ respectively. This is likely an underestimate of the number of compact binary BHs, because it considers only those formed from isolated evolution. The common envelope phase that leads to such compact binaries may be triggered by dynamical interactions, therefore increasing the fraction $f_{\rm CBBH}$ \citep{kumamoto2019,dicarlo2019}. The fraction of single black holes is instead $f_{\rm BH} \simeq 0.028$, consistent with an evolving population of stars that follows a \citet{kro01} mass function between 0.8 and 150 solar masses.

Finally, we can obtain the encounter rate averaged over the star cluster mass function, which follows a power-law of $\beta = -2$, as observed in massive clusters in the Galactic disk and starburst galaxies \citep{por10}:
\begin{equation}
	 \Gamma^{2+1}_\mathrm{enc}  = A_\mathrm{cl} \int^{10^6 \msun}_{10^2 \msun} \Gamma^{2+1}_\mathrm{enc}(M_\mathrm{cl})  M_\mathrm{cl}^{-2} d M_\mathrm{cl}
\end{equation}
where $A_\mathrm{cl}$ is the normalization factor so that the cluster mass function normalizes to 1. The averaged encounter rate is then
\begin{equation}
	 \Gamma^{2+1}_\mathrm{enc} \simeq 10^{-3} \myr^{-1} \left(\frac{f_{\rm CBBH}}{10^{-3}}\right) \left(\frac{f_{\rm BH}}{0.02}\right)
\end{equation}
Given the short lives of OCs (${\sim}300\myr$, \citealt{por10}), it seems that OCs might only experience a few such binary-single encounters, if any. This low encounter rate justifies our assumption of considering only the effect of a single encounter.

On the other hand, the rate of binary-binary encounters in OCs dominates over that of binary-single encounters due to the abundance of wide stellar binaries \citep[e.g.][]{leigh2011,leigh13b,geller15}. While we limit ourselves to simulating binary-single encounters, we do not expect the outcome of binary-binary encounters to be statistically different. The reason is that wide binaries in OCs have typically a semimajor axis much larger than $10\au$, which is $10^2$--$10^3$ larger then the hard binaries of our sample \citep{ragh10}. Such hard-soft binary encounters tend to quickly eject one of the wide binary members, and subsequently continue the evolution as a three-body encounter. 

The binary-binary encounter rate is $\Gamma^{2+2}_\mathrm{enc}$ and can be calculated as 
\begin{equation}\label{eq:encbinbin}
	\Gamma^{2+2}_\mathrm{enc} = f_{\rm bin,a} f_{\rm bin,b} \, N n \, \frac{8\pi G m_\mathrm{tot} R_\mathrm{enc}}{\sigma_\infty}
\end{equation}

Assuming the typical size of wide binaries, ($R_\mathrm{enc} \equiv a_\mathrm{WB} = 30 \au$, \citealt{ragh10}), $f_{\rm bin,a} \equiv f_{\rm CBBH} = 10^{-3}$, and a wide binary fraction of $f_{\rm bin,b} \equiv f_{\rm WB} = 0.5$, Equation~\ref{eq:encbinbin} leads to a much higher encounter rate of $\Gamma^{2+2}_\mathrm{enc} \simeq 40 \myr^{-1}$. However, only close passages between the compact binary BH with a binary member can lead to a meaningful encounter: given its compactness, the binary BH can simply pass through the wide binary without really interacting as a 2-body object. Therefore, to estimate the binary BH-wide binary encounter rate we set $R_\mathrm{enc} \equiv 2 \ainn$, and double the rate to take into account that each wide binary is composed of two objects. With this, the 2+2 encounter rate becomes $\Gamma^{2+2}_\mathrm{enc} \simeq 0.2 \myr^{-1}$. In this assumptions we ignore that some of these encounters may end with the wide binary companion bound to the compact binary, forming a stable hierarchical triple. This triple would be too wide to affect the evolution of the compact binary via the von~Zeipel-Kozai-Lidov mechanism \citep{zeipel1910,lid62,koz62}, and would likely be disrupted via a subsequent interaction on a very short timescale.

Note that only encounters with wide binary BHs, rather than stellar wide binaries, can lead to a desirable outcome, that is tilting of the orbital plane of the binary BH. Most encounters with stellar objects will involve low-mass main sequence stars, because by the time the primordial binary has become a compact binary BH, massive stars have already collapsed into BHs. Such low-mass stars will have a limited impact on the more massive binary BH. To significantly affect the binary BH, the velocity kick from the passing star needs to be comparable to the orbital velocity of the binary BH, $v_\mathrm{bin} = \sqrt{G m_\mathrm{bin}/\ainn}$. From conservation of linear momentum, this requires the star to approach one of the binary members by at least ${\sim} a_\mathrm{bin} (m_\mathrm{star}+m_\mathrm{BH}) / m_\mathrm{bin}$. This distance is about $0.019 \au$ for a $1 \msun$ star encountering a $40\msun$ binary BH; this distance is dangerously close to the stellar tidal disruption radius $R_\mathrm{star} (m_\mathrm{star}/m_\mathrm{BH})^{1/3} \sim 0.012 \au$. Hence most of the encounters between compact binary BHs and wide stellar binaries will result in either little impact on the binary BH or stellar tidal disruptions. We can then correct the binary-binary encounter rate by taking into account only encounters with wide binary BHs, whose fraction we can estimate as $f_{\rm BH} f_{\rm WB} \simeq 0.014$. In this approximation we have neglected stellar binary interactions, which are unlikely to occur in wide binaries. The total encounter rate is therefore $\Gamma_\mathrm{enc} = \Gamma^{1+2}_\mathrm{enc} + \Gamma^{2+2}_\mathrm{enc} = 6.6 \times 10^{-3} \myr^{-1}$.

This model assumes that most of the encounters will occur in the core, which may not be always correct. \citet{barrera2020} find that the total integrated (i.e., over the entire volume of the cluster) encounter rate is underestimated by a factor of ${\sim}5$ when compared to the core rate, as confirmed via N-body simulations. \citet{barrera2020} found that of order 50\% of all interactions occur in the core, with a non-negligible additional contribution coming from interactions occurring just outside the core, and then drifting into the core due to mass segregation.  Hence, the correction factor from the integrated rate calculation can simply be multiplied by the total core rate, in order to obtain a total rate for the entire cluster.

It is worth reminding that this cross-section estimate is rather simplistic because it neglects the role of global stellar cluster processes, such as core collapse, dynamical friction and mass segregation, which tend to increase the frequency of encounters.
The encounter rates outlined here are to be intended as a lower limit estimate. In other words, an enhanced rate of three-body encounters are an unavoidable consequence of the gravothermal instability of self-gravitating systems, and can accelerate the disruption of star clusters \citep{leigh2014,leigh2016}.

\section{Mergers properties}\label{sec:disc}

\subsection{Local merger rate}\label{sec:locmerg}
To calculate the local merger rate density, we adopt the same approach of \citet{kumamoto2020}. Particularly, we use their equation (16) to express the local merger rate density:
\begin{equation}\label{eq:mrate}
	\Gamma^\mathrm{loc}_\mathrm{gw} = \frac{9\times 10^{-4}}{4\msun \ln{10}} \int dZ \int dt_\mathrm{L} D(Z,t_\mathrm{gw}=t_\mathrm{L}) \frac{d\Psi(Z,t_\mathrm{L})}{dZ}
\end{equation}
where $D(Z,t_\mathrm{gw})$ is the merger rate of binary BHs originated from one cluster,  $t_\mathrm{L}$ is the lookback time, and $\Psi(Z,t_\mathrm{L})$ is the comoving formation rate density of stars.

We calculate $D(Z,t_\mathrm{gw})$ from the simulations, and include $\Psi(Z,t_\mathrm{L})$ as estimated by \citet{chruslinska2019}. We express $D(Z,t_\mathrm{gw})$ as the product of the binary encounter rate $\Gamma_\mathrm{enc}(Z)$, and the density distribution of delay times $B(Z, t_\mathrm{gw})$:
\begin{equation}\label{eq:dfunc}
	D(Z,t_\mathrm{gw}) = \Gamma_\mathrm{enc}(Z)\, B(Z,t_\mathrm{gw})
\end{equation}
Here the encounter rate depends on the metallicity of the parent cluster through the compact binary BH fraction $f_{\rm CBBH}$, as estimated as in Section~\ref{sec:encrates} ($\Gamma_\mathrm{enc} \simeq 0.0005, 0.0066$ and $0.0143 \myr^{-1}$ for $Z=1, 0.1$ and $0.001\zsun$, respectively). The density distribution of delay times $P(Z,t_\mathrm{gw})$ is obtained from the OC three-body simulations.

We consider three metallicity ranges: $Z=0.00632$--$0.1$, $0.000632$--$0.00632$, and $0.000001$--$0.000632$ for the simulations at $1$, $0.1$, and $0.01 \zsun$, respectively (i.e. equally spaced in logarithmic scale). Ultimately, the expression for the local merger rate reads as:
\begin{equation}
	\Gamma^\mathrm{loc}_\mathrm{gw} = \frac{9\times 10^{-4}}{4\msun \ln{10}} \sum_{z=1,2,3} \int  D_{z}(t_\mathrm{gw}=t_\mathrm{L}) \Psi_{z}(t_\mathrm{L})dt_\mathrm{L}
\end{equation}
where the summation is over the three ranges of metallicities, corresponding to the simulations at $Z = 1, 0.1$ and $0.01\zsun$, and $\Psi_{z}(t_\mathrm{L})$ is the star formation rate density $d\Psi(Z,t_\mathrm{L})/dZ$ integrated over the three ranges (see equations 31--35 from \citealt{kumamoto2020}).

We obtain a local merger rate of 
\begin{equation}\nonumber
	\Gamma^\mathrm{loc}_\mathrm{gw} \simeq 6.6 \yr^{-1} \,\rm Gpc^{-3}
\end{equation}

Despite the higher abundance of compact binary BHs at $Z = 0.01 \zsun$ ($f_{\rm CBBH} \simeq 2\times 10^{-3}$), their contribution to the local merger rate is only ${\approx}2 \yr^{-1} \,\rm Gpc^{-3}$. The main reason is that most low-metallicity BHs are born at high redshift \citep[see fig. 6][]{chruslinska2019}, but they have a very short delay time (Figure~\ref{fig:tgw}). Hence, the contribution to our estimated merger rate comes also from solar and moderately sub-solar metallicity binary BHs.

While our estimated merger rate lies at the lower limit inferred by GWTC-2 \citep{gwtc2020b}, it only applies to the subset of BH-BH mergers undergoing the pathway considered here. Our estimate indicates that such binaries may be already contributing to the current detection sample, and that they will likely emerge from the data after a few hundred detections. 

\subsection{\texorpdfstring{$\chi_\mathrm{eff}$ and $\chi_\mathrm{p}$ distributions}{Xeff and Xp distributions}}\label{sec:xeff}
The information on BH spin during the merger is encoded into two parameters, the effective spin parameter $\chi_\mathrm{eff}$ and the effective precession parameter $\chi_\mathrm{p}$. Both parameters describe the orientation of the spin vector with respect to the binary orbit: the effective spin $\chi_\mathrm{eff}$ relates to the spin component parallel to the orbital angular momentum vector, while the effective precession $\chi_\mathrm{p}$ relates to the orbital precession caused by the in-plane spin component.

Given the dimensionless spin $\chi$ and the spin obliquity $\theta$ (i.e. the angle between the spin vector and the orbital angular momentum vector), we can express $\chi_\mathrm{eff}$ and $\chi_\mathrm{p}$ as:
\begin{equation}
\begin{split}
\chi_\mathrm{eff} & = \frac{m_1 \chi_1 \cos{\theta_1} + m_2 \chi_2 \cos{\theta_2} }{ m_1 + m_2} = \\ & = \frac{(m_1 \chi_1 + m_2 \chi_2)}{ m_1 + m_2} \cos{\Delta i}
\end{split}
\end{equation}

\begin{align}
\begin{split}
\chi_\mathrm{p} & = \max{\left[ \chi_1 \sin{\theta_1}, \chi_2 \sin{\theta_2}\frac{ 4q + 3}{4 + 3q}q \right]} = \\
& = \max{\left[ \chi_1, \chi_2 \frac{ 4q + 3}{4 + 3q}q \right]} \sin{\Delta i} 
\end{split}
\end{align}
where the indices $_1$ and $_2$ refer to the primary and secondary BHs, so that $m_1 > m_2$, and $q = m_2 / m_1 \leq 1 $ is the mass ratio.  
The last identity takes into account our initial setup, where both spins are initially aligned and $\theta_1 = \theta_2 = \Delta{i}$. 

The dimensionless spin of BHs at birth is largely uncertain. Its precise value depends on the interplay between winds and tidal interactions of the progenitor binary, and on the physics of angular momentum transport during the last stages of core collapse \citep[e.g.][]{quin2018,bavera2020}. \rev{These processes are largely uncertain, and they are commonly described by parametrized models. To avoid introducing further model degeneracies, we adopt 3 different phenomenological models for the dimensionless spin.} In the \textsc{maximum} model, we assume that both BHs in the original binaries are born maximally spinning, i.e. $\chi_1 \equiv \chi_2 \equiv 1$. In the \textsc{uniform} model, the spin of the BHs is instead randomly drawn from a uniform distribution between 0 and 1. In the \textsc{beta} model, the spin is sampled from a beta distribution with scale parameters $(\alpha,\beta) = (2.3, 4)$. The beta distribution peaks at about $\chi \approx 0.2$ with a long tail at $0.3$--$0.7$. The spin of the initially isolated BH can affect the effective spin distributions of exchanged binaries; it is assumed to be zero in all three models.

\rev{In physical terms, the \textsc{beta} model is roughly consistent with a super-efficient angular momentum transport mechanism via the Tayler-Spruit magnetic dynamo \citep{spruit1999,spruit2002,fuller2019,fuller2019a}. In this model, all black holes are born with negligible spin ($\chi \simeq 10^{-2}$), unless their Wolf-Rayet progenitors are spin-up by tidal interactions.}

\begin{figure}
	\centering
	\includegraphics[width=1\columnwidth]{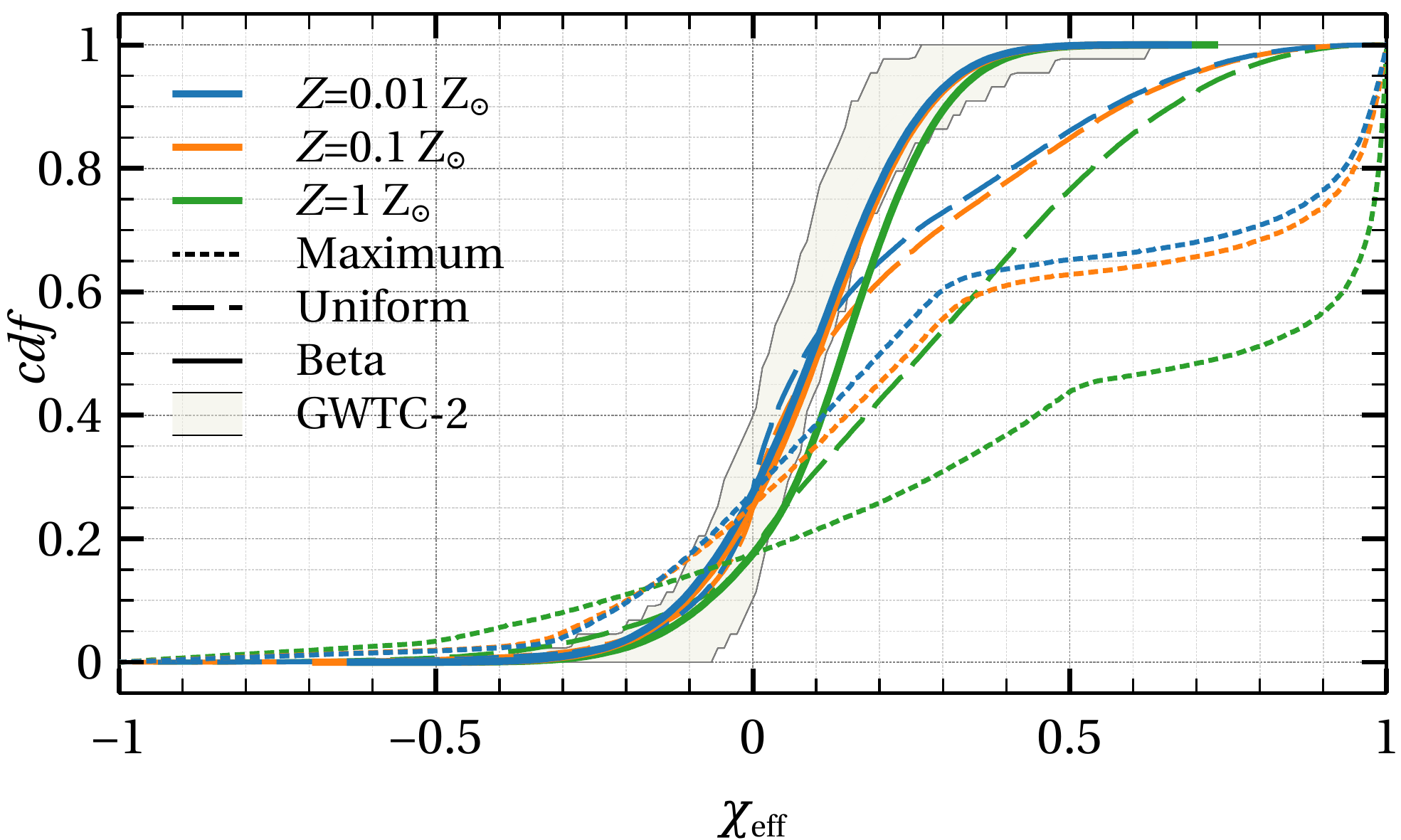}
	\includegraphics[width=1\columnwidth]{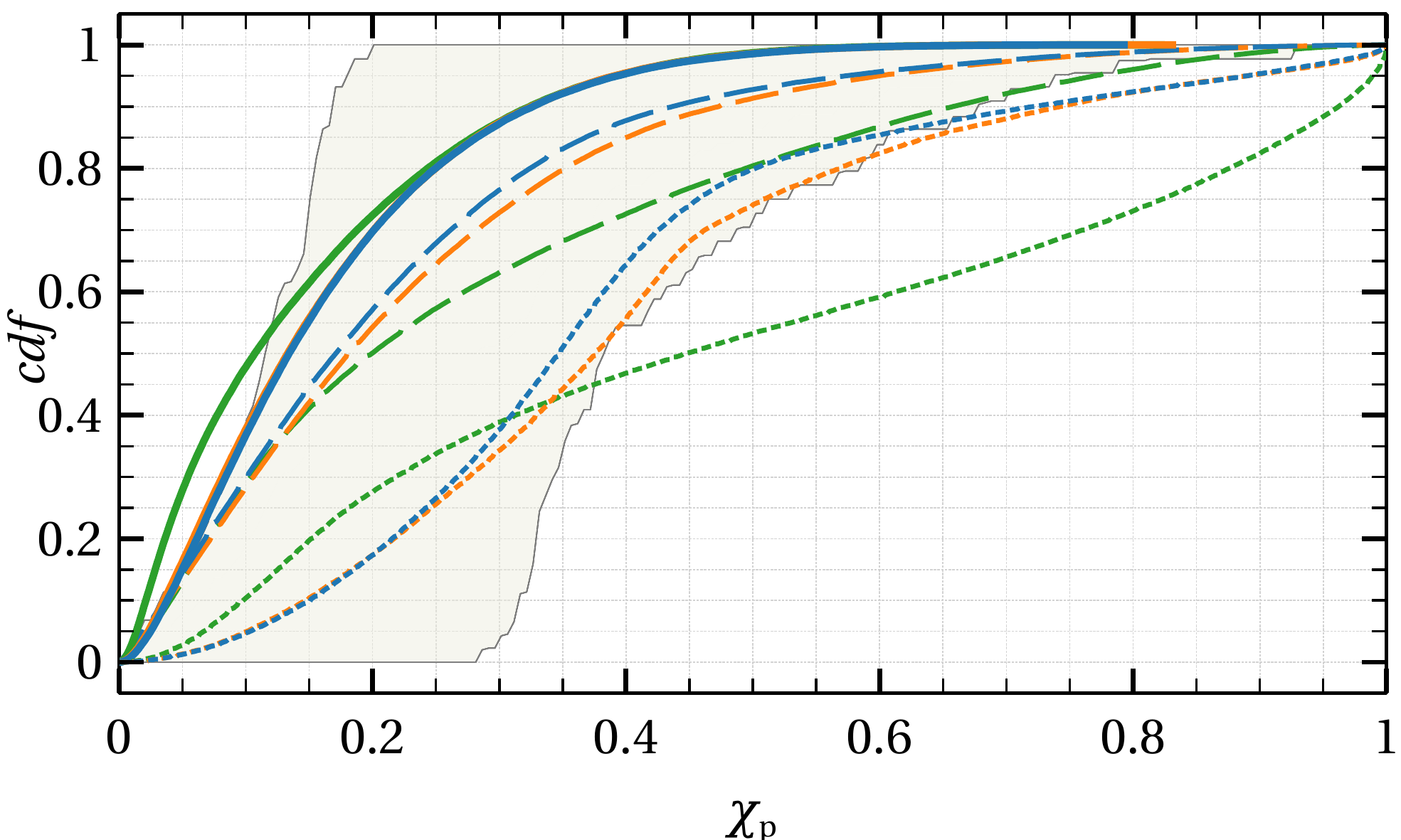}
	\caption{Cumulative distributions of $\chieff$ (top) and $\chip$ (bottom) for each simulated set. Green lines: $Z=1\zsun$. Orange lines: $Z=0.1\zsun$. Blue lines: $Z=0.01\zsun$. The OC and GC distributions overlap, so we show only the OC ones. The line style denotes the model for the natal dimensionless spin $\chi$. Dotted lines: maximally spinning BHs ($\chi_1 \equiv \chi_2 \equiv 1$). Dot-dashed line: $\chi_1$, $\chi_2$ drawn from a uniform distribution in (0,1). Solid lines: $\chi_1$, $\chi_2$ sampled from a beta distribution with $(\alpha,\beta) = (1.3, 4)$. In all three models the isolated BHs are non-spinning $\chi_3 \equiv 0$. The gray shaded area shows the 90\% confidence interval of the distribution reconstructed from GWTC-2 \citep[default spin model of][]{gwtc2020b}.}
	\label{fig:Xcum}
\end{figure}

Figure~\ref{fig:Xcum} shows the cumulative distribution of $\chieff$ and $\chip$ for each simulated set. Because the outcome distributions are the same regardless of the initial velocity dispersion of the three-body encounter, we omit plotting the curves from the GC simulations.

In the \textsc{maximum} model, the $\chieff$ distributions are characterized by a peak at $\chieff \sim 1$, stronger at high metallicity. This peak is mainly composed of original binaries that experienced only weak encounters with moderate tilting. Hence, the peak at $\chieff\sim 1$ simply traces the distribution of $\chieff$ before the encounter, which is identical to $1$ in the \textsc{maximum} model. 
The slope between $\chieff \sim -0.5,0.5$ is composed of exchanged binaries, wherein the primary BH is the non-spinning, exchanged one. Overall, the slope in the \textsc{maximum} model is too shallow the match the one inferred from the observations.

Going from the \textsc{maximum} to the \textsc{beta} model, the average dimensionless spin of the BHs decreases, and the cumulative distributions of $\chieff$ becomes more peaked at zero. Particularly, the distribution at $Z=0.1$ and $0.01 \zsun$ for the \textsc{beta} model matches well the inferred distribution from GWTC-2.

The $\chip$ cumulative distribution has a similar trend, with the \textsc{beta} model more peaked at zero, and the \textsc{maximum} model favoring larger $\chip$. Overall, the \textsc{uniform} model matches better the $\chip$ constraints from the GWTC-2 data.

\begin{figure}
	\centering
	\includegraphics[width=1\columnwidth]{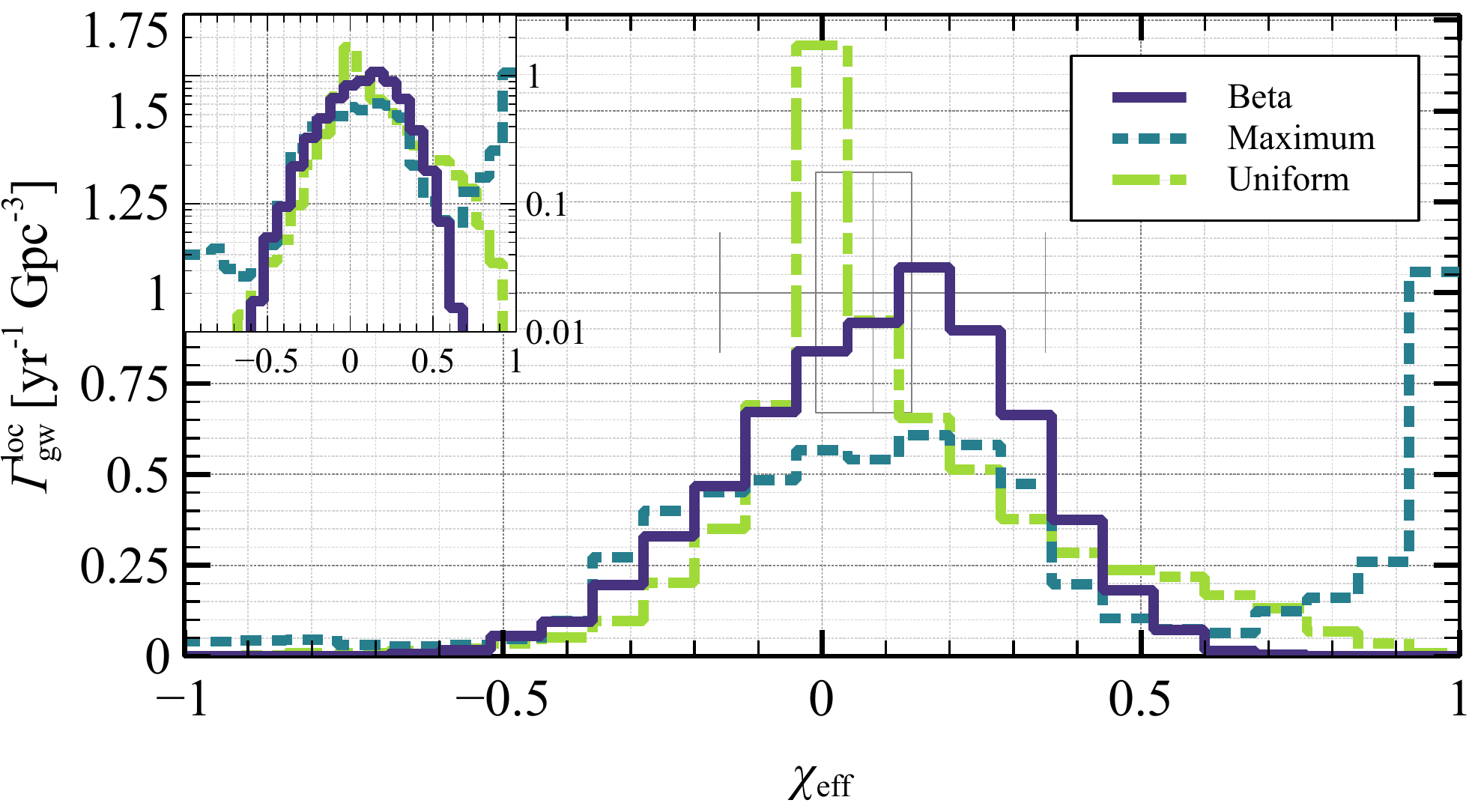}
	\includegraphics[width=1\columnwidth]{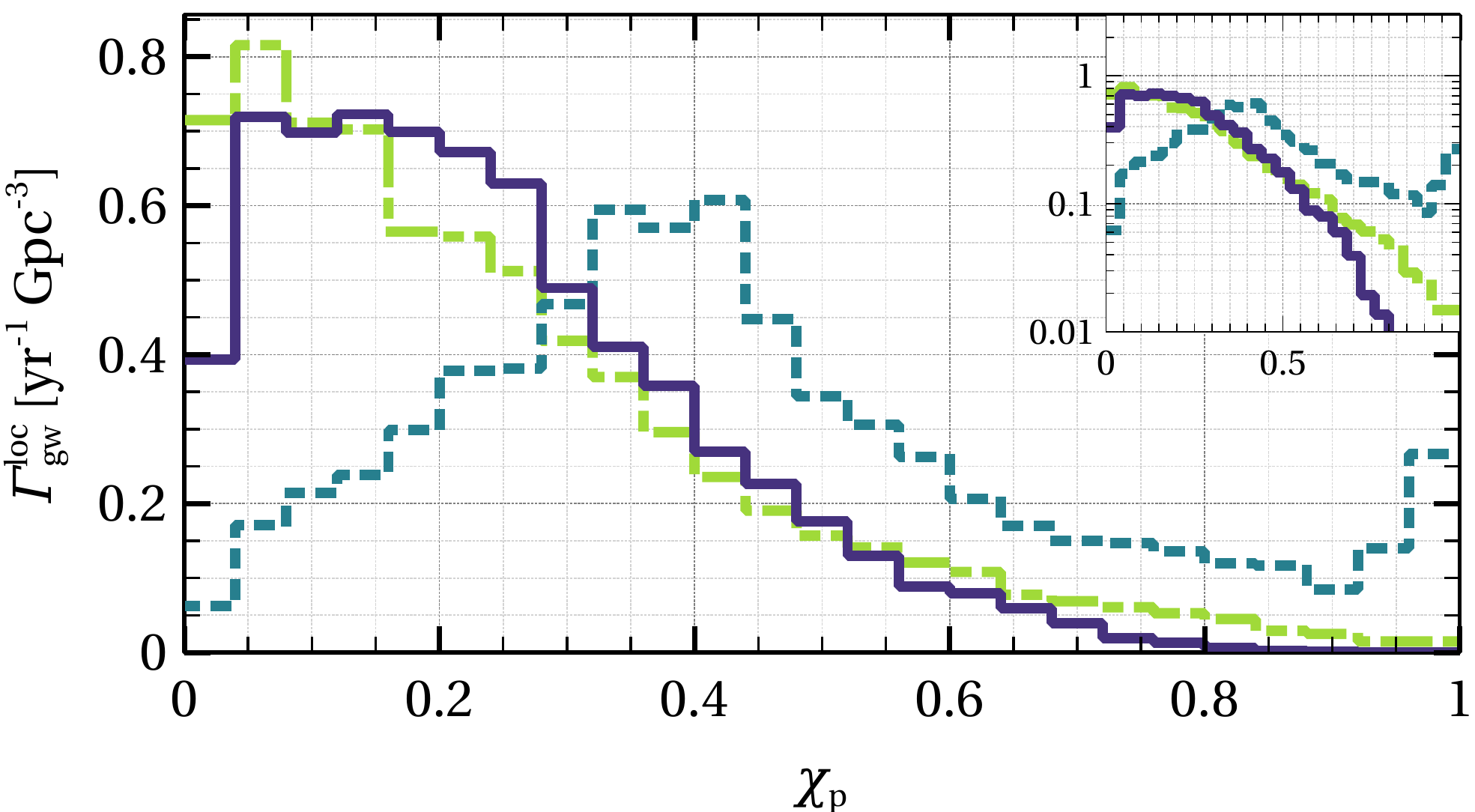}
	\caption{Merger rate density as a function of the binary $\chieff$ (top) and $\chip$ (bottom), calculated as in Section~\ref{sec:locmerg}. Each curve corresponds to a different model for the natal BH spin. The \textsc{beta} model (dark violet, solid line) favors low dimensionless spins (${\approx}0.2$). The \textsc{maximum} model (teal, dashed line) assumes maximally spinning BHs. The \textsc{uniform} model (bright green, dot-dashed line) draws the spin from a uniform distribution in (0,1). \rev{The box plot in the top panel shows the 1.5 inter-quartile range using the raw GWTC-2 data.}}
	\label{fig:Xrates}
\end{figure}

In addition to displaying the distribution of $\chieff$ and $\chip$ for each metallicity, we compute the local merger rate for different bins of $\chieff$ and $\chip$. Each domain range is divided in 25 uniform bins, and the procedure to calculate the local merger rate is repeated for each binary subset. Figure~\ref{fig:Xrates} shows the obtained local merger rate as a function of $\chieff$ and $\chip$. The end result can be thought of as a combination of the distributions in Figure~\ref{fig:Xcum}, weighted by merger rate per metallicity range, plus second order effects from the correlations between effective spin and delay time.

The \textsc{beta} spin model best matches the current observational data. The $\chieff$ distribution is compatible to the observed one, with a peak slightly above 0 and a tail at negative $\chieff$. The $\chip$ distribution has a broad peak at ${\approx}0.2$, in reasonable agreement with the GWTC-2 data \citep[see fig. 9 of][]{gwtc2020b}. 

\rev{In contrast, the \textsc{maximum} model predicts a peak at $\chieff \sim 1$, which is not present in the data. The $\chieff \sim 1$ peak is constituted by original binaries that were only weakly perturbed by the encounter, and consequently remained largely aligned. 
The \textsc{uniform} model strongly favors zero values of $\chip$ and $\chieff$, which are also disfavored by present data.

Regardless of the spin models, the $\chieff$ distribution is skewed towards positive values. This is in net contrast with the distribution predicted by the dynamical assembly scenario, which is symmetric around zero \citep[e.g.][]{rodriguez2019}. The reason for this comes from the distribution of tilt angles $\Delta i$ (Figure~\ref{fig:dangletot}), which is anisotropic even for exchanged binaries. The correlation between the initial and final orbital orientation is not entirely erased by the encounter, and thus spin-orbit alignment is remains favored.

Interestingly, a distribution of $\chieff$ similar to our \textsc{beta} spin model was obtained by \citet{belczynski2020} from isolated binary evolution models. In their scenario, the misalignment is caused by the natal BH kicks, which at high magnitude velocity can even flip the binary, leading to negative $\chieff$. The origin of this model degeneracy arises from the following reason. In a broad sense, very high natal kicks can have the same effect of a ``prompt'' three-body interaction, which in fact can be modeled in the impulse approximation \citep[e.g.][]{micheaely2019}. However, many three-body interactions are constituted by repeated excursions and encounters, whose effect tends to isotropize the  binary orientation. In our case, the initial binaries are already very compact, so that most of them undergo only a single strong interaction prior to ejection. Therefore, in our case, three-body encounters can mimic the effect of strong BH natal kicks.

On the other hand, models of isolated binary evolution with low natal kicks are not able to produce binaries with anti-aligned spins and consequently GW events with negative $\chieff$ \citep[e.g.][]{bavera2020,callister2020}.
}


\section{Caveats}

We remind here the assumptions we made along this work. First, we have considered only compact binary BHs undergoing a single three-body encounter. This was justified in Section~\ref{sec:encrates} and Section~\ref{sec:res}, by noting that the encounter rate of such binaries is small compared to cluster lifetimes and GW coalescence times. Moreover, the binaries get ejected from most stellar clusters due to the high recoil kicks from the encounters, preventing further encounters. 

Another assumption of our work is that the BH spins are initially aligned with the orbit. This assumption is reflected in the choice of the initial conditions: all our binary BHs come from post-common envelope evolution, and form a close Wolf-Rayet binary before collapsing into BHs. Our binary sample has a median period less than 1 day, so that tidal forces can efficiently align the spin of the progenitor stars \citep{kushnir2016,hotokezaka2017,piran2020}. In general, this is not true for longer binary periods and stars that underwent significant mass transfer; however this depends on the tidal efficiency and the angular-momentum transport within the stellar interiors, which are highly uncertain \citep{stegmann2020}.

Finally, we neglected BH natal kicks that might tilt the binary plane and the stellar spins. In our case, the kick magnitude should be ${>}600 \kms$ to affect the binary angular momentum, which is unlikely \citep{mandel2016a,mirabel2016,wysocki2018}.

\section{Conclusions}

We investigated the spin parameter distributions of post-common-envelope binaries that undergo dynamical encounters in stellar clusters. This binary formation pathway was identified by \citet{kumamoto2019} and \citet{dicarlo2020b} in numerical simulations of young star clusters. Both studies showed that this pathway contributes significantly to the binary BH merger rate, especially at low metallicity.

We assume that binary BHs emerge from this evolutionary pathway with spins aligned with the orbital angular momentum, and subsequently undergo a three-body encounter with an isolated BH. The encounter can tilt the orbital plane of the binary, resulting in spin-orbit misalignment. The orbital tilt can therefore lead to GW signals with non-zero $\chip$ and negative $\chieff$, even if the BH spins remain aligned with each other. We consider only one single encounter, as justified by the high recoil velocity of the binary (Figure~\ref{fig:vescplot}) and by the low encounter rate of post-common-envelope binaries (Section~\ref{sec:encrates}).

We infer the distribution of orbital tilt angles $\Delta i$ after a single encounter by means of direct N-body integrations. We model the encounter with the highly-accurate few-body code \textsc{tsunami}, including post-Newtonian corrections to the equations of motion. 
Our binary initial conditions are drawn from an updated version of the population stellar synthesis code \textsc{bse} \citet[][see also \citealt{tanikawa2020b}]{tanikawa2020c}, considering three different metallicities: $Z=1,0.1$ and $0.01 \zsun$. We select only the binaries that survive common envelope evolution as Wolf-Rayet stars and that are spun-up by tidal interactions. 
The mass of the third BH is consistently selected from single stellar population synthesis. For the encounter properties, we consider both low velocity dispersion environments (corresponding to open clusters) and high velocity dispersion environments (corresponding to globular and young massive star clusters). We show that the encounter outcome is the same regardless of the velocity dispersion, due to the compactness of the binaries. 

\rev{While it is generally assumed that dynamical exchanges in star clusters result in an isotropic orbit-spin misalignment, we show that it is not the case for compact post-common envelope binaries, which are limited to a single strong interaction. The orbital tilt angle distribution (Figure~\ref{fig:dangletot}) for exchanged binaries is not entirely flat, but still favors mild spin-orbit alignment. This results in a $\chieff$ distribution skewed towards positive values, in contrast to the symmetric distribution predicted by the dynamical assembly scenario.}

We estimate the local merger rate considering the cosmic star formation rate density at different metallicities, taking into account the delay time of BH mergers and the binary encounter rates. We obtain a \rev{lower limit for the} local merger rate of $\Gamma^\mathrm{loc}_\mathrm{gw} \simeq 6.6 \yr^{-1} \,\rm Gpc^{-3}$, which shows that this pathway might be contributing to the events detected so far.

We also estimate the differential merger rate for the effective $\chieff$ and precession $\chip$ spin parameters.
Because the value of natal BH spins $\chi$ is subject to numerous uncertainties, we test three different phenomenological models. 

Assuming low dimensionless spins $\chi \lesssim 0.2$ in binaries and non-spinning isolated BHs, our scenario \rev{qualitatively reproduces the distributions of $\chip$ and $\chieff$ inferred from GWTC-2. In particular, this model can explain the peak at positive $\chieff \sim 0.1$  with a tail at negative $\chieff$ and the broad peak at $\chip\sim 0.2$ in the precession spin parameter}.

\section*{Acknowledgements}
We thank the referee for their constructive review of this manuscript. This work was supported by JSPS KAKENHI Grant Number 17H06360, 19H01933, and 19K03907, MEXT as “Program for Promoting Researches on the Supercomputer Fugaku” (Toward a unified view of the universe: from large scale structures to planets, Revealing the formation history of the universe with large-scale simulations and astronomical big data), and The University of Tokyo Excellent Young Researcher Program. The simulations were run on the CfCA Calculation Server at NAOJ. N.W.C.L. gratefully acknowledges support from the Chilean government via Fondecyt Iniciaci\'on Grant \#11180005. 

\section*{Data Availability}

The \textsc{tsunami} code, the initial conditions and the simulation data underlying this article will be shared on reasonable request to the corresponding author.



\bibliographystyle{mnras}
\bibliography{totalms,additionalref} 




\appendix




\bsp	
\label{lastpage}
\end{document}